\documentclass[aps,prl,preprint,groupedaddress]{revtex4-1}

\usepackage{graphicx}
\usepackage{bm}
\usepackage{amsmath}
\begin{document}


\title{Nonlinear system synchronization to sum signals of multiple chaotic systems}


\noaffiliation

\noaffiliation


\noaffiliation

\noaffiliation

\noaffiliation

\author{Robson Vieira, Weliton S. Martins, Sergio Barreiro, Rafael A. de Oliveira, Martine Chevrollier, Marcos Ori\'a}
\affiliation{Universidade Federal Rural de Pernambuco - UACSA\\
Cabo de Santo Agostinho - PE - Brazil}
\noaffiliation


\date{\today}

\begin{abstract}

Coupling of chaotic oscillators has evidenced conditions where synchronization is possible, therefore a nonlinear system can be driven to a particular state through input from a similar oscillator. Here we expand this concept of control of the state of a nonlinear system by showing that it is possible to induce it to follow a \textit{linear} superposition of signals from multiple equivalent systems, using only partial information from them, through one- or more variable-signal. Moreover, we show that the larger the number of trajectories added to the input signal, the better the convergence of the system trajectory to the sum input.

\end{abstract}


\maketitle

\section{Introduction}
 
Nonlinear systems, such as Lorenz \cite{lorenz1963}, R\"{o}ssler \cite{rossler1976}, or Gauthier-Bienfang \cite{gauthier1996intermittent} systems, have been numerically shown to converge to a particular solution by adding into the system information on a particular trajectory, either through substitution of variables \cite{pecora1990synchronization} for the synchronization of chaotic oscillators, or by adding a linear term \cite{gauthier1996intermittent} to the original system. Here we show that it is possible to lead a nonlinear system to a new defined trajectory by driving it with a linear combination of signals generated from N independent equivalent systems. Moreover, the system trajectory better mimics that of the sum signal when the number of added trajectories increases.

In order to investigate this behavior of synchronization of a nonlinear system with a sum signal  obtained from two or more drives, we prepare $N$ similar systems with slightly mismatched parameters, oscillating independently from one another. Each drive $\textbf{X}_{n}$ is a particular solution of the free-running system equation
\begin{equation}
\label{eq1}
\mathbf{\dot X}_{n}= \mathbf{F}\left( \mathbf{X}_{n}\right), 
\end{equation} 
where $\mathbf{X}_{n}^{T} = (x_{n}, y_{n}, z_{n})$ and $\mathbf{F}$ is the vector field describing the flux of the system. Information from the drives trajectories is sent to the nonlinear receiving oscillator by way of a sum signal, as described by the modified equation,
\begin{equation}
\label{eq2}
\mathbf{\dot X} = \mathbf{F}\left( \mathbf{X} \right) + \mathbf{K} \sum_{n=1}^{N} c_{n} \left(\mathbf{X}_{n} - \mathbf{X}    \right),  
\end{equation} 
where $\mathbf{X}$ represents the receiving oscillator in $\mathbf{\Re^{3}}$, with $\mathbf{X}^{T} = (x, y, z)$. $\mathbf{K}$ is a $m \times m$ matrix allowing for the coupling of the $\mathbf{X}$ components and  $c_{n}$ represents the strength of the coupling for each solution, $\mathbf{X}_n$. Here we numerically show that
\begin{equation}
\label{eq3}
\mathbf{X} \longrightarrow \mathbf{X_s} = \dfrac{\sum_{n=1}^{N} c_{n} \mathbf{X}_n }{ \sum_{n=1}^{N} c_{n}},
\end{equation}
i.e., for large values of $N$ and positive $c_{n}$, the trajectories of the chaotic response oscillator  $\mathbf{X}$  converge to $\mathbf{X_{s}}$, with $\mathbf{X_s}^{T} = (x_s, y_s, z_s)$. 

To demonstrate statement (\ref{eq3}) we generated solutions $\mathbf{X}_n$ $(n = 1, 2, 3,...)$ of Eq. (\ref{eq1}) starting from arbitrary initial conditions for each solution and allowing small mismatches between drives, i.e., the parameters defining $\mathbf{F}$ may be slightly modified  (by a few percent) in order to check for the robustness of the technique. The generated drive solutions are then linearly combined and introduced in Eq.(\ref{eq2}). We establish which variables are coupled through the choice of $\mathbf{K}$ components. We applied this technique to a few nonlinear systems such as Lorenz, R\"{o}ssler, and Gauthier-Bienfang systems. The results are discussed below.

\section{Results}
\label{results}

We present here in some detail our syncronization technique applied to a Gauthier-Bienfang system. The flow $\mathbf{F}$ in equations (\ref{eq1}) and (\ref{eq2}) takes the form
\begin{equation}
\label{eq4}
\mathbf{F} = 
\begin{bmatrix}
  x_n - g[x_n - y_n] \\
  g[x_n - y_n] - z_n \\
  y_n - C z_n
\end{bmatrix},
\end{equation}
where $g[\chi] = \chi/B + D \left[\exp(\alpha \chi) - \exp(-\alpha \chi)\right]$, and $\alpha$, $A$, $B$, $C$, and $D$ are positive constants in $\mathbf{\Re}$ [3]. We initially analyzed the system with two drives, with  $\mathbf{X}_1$ and $\mathbf{X}_2$ their respective trajectories, solutions of equation (\ref{eq1}). For the sake of simplicity we consider $c_1 = c_2= 1.0$.  We solved numerically equation (\ref{eq2}) for the particular flow $\mathbf{F}$ described in eq. (\ref{eq4}), using $K_{11} = K_{22} = 1.0$, and $K_{ij} = 0$, for all the other components.

\begin{figure}[htb]
\centering 
 \includegraphics[width = 8.0 cm]{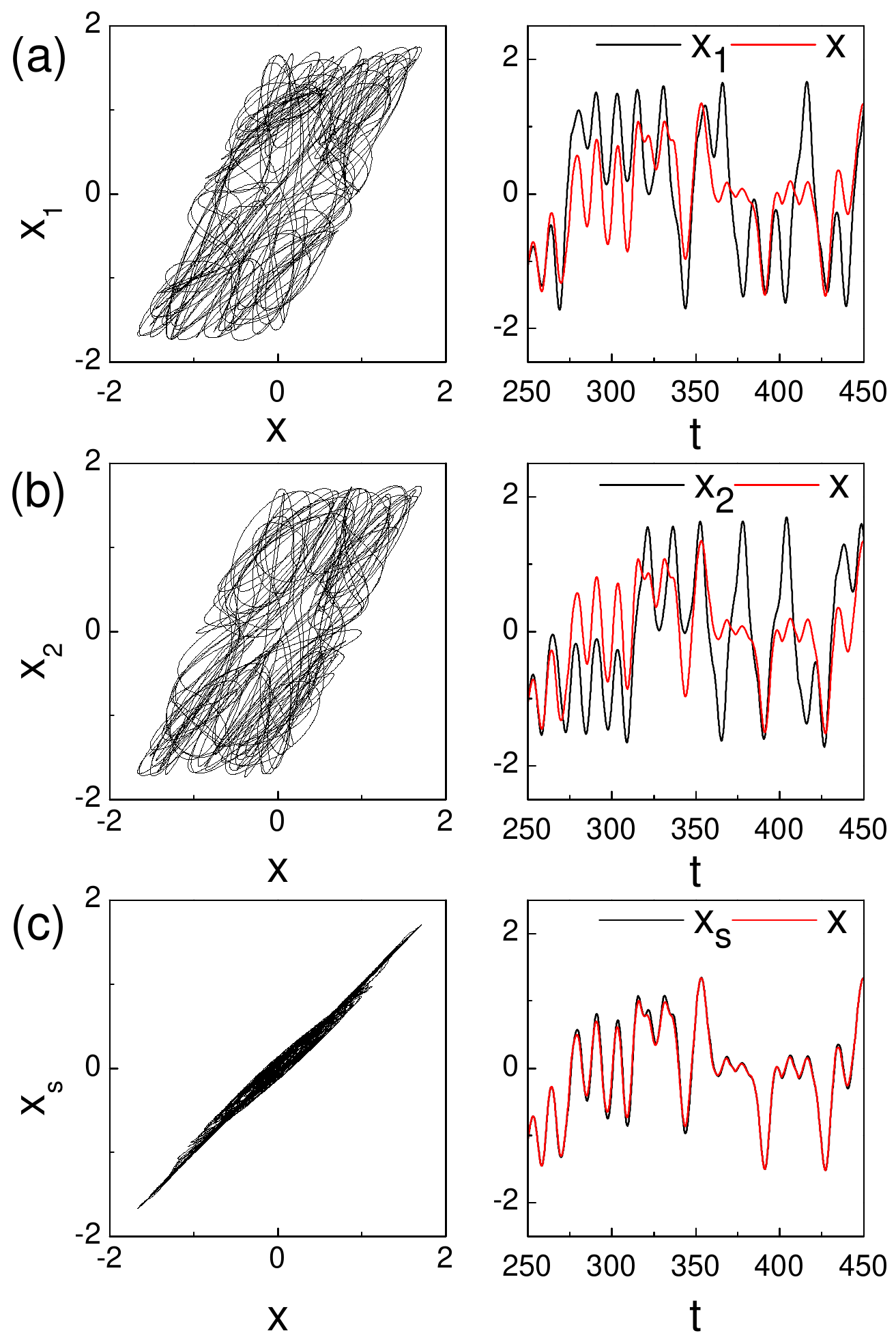}
 \caption{\label{fig1} Output of the Gauthier-Bienfang nonlinear system with addition of a linear combination of 2-solution coupling. (a) $x_1 \times x$; (b) $x_2 \times x$; (c) $x_s = \frac{(x_1 +x_2)}{2} \times x$. Left: synchronization analysis; Right: temporal series.}
\end{figure}

\begin{figure}[t]
\centering 
 \includegraphics[width = 8.0 cm]{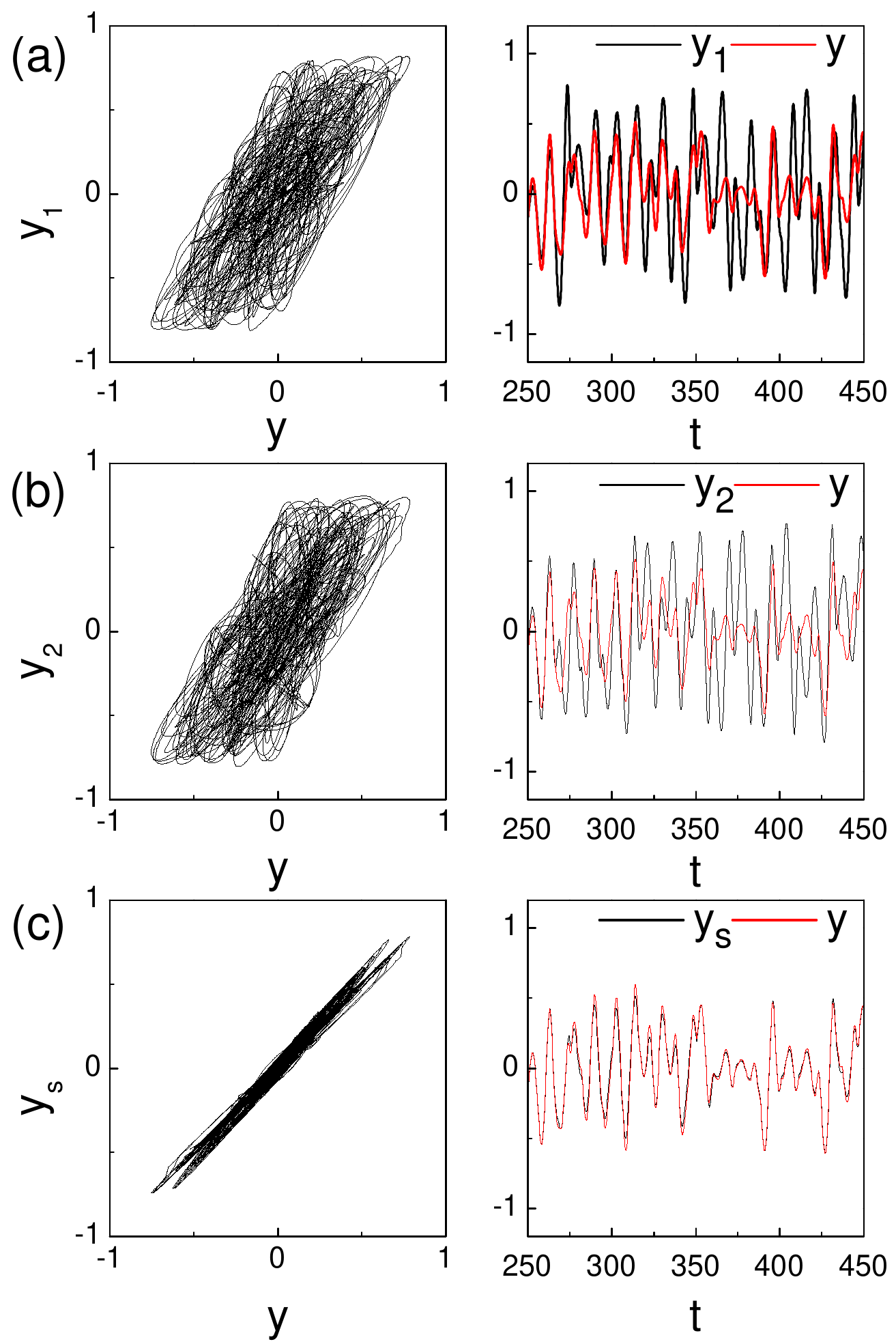}
 \caption{\label{fig2} Output of the Gauthier-Bienfang nonlinear system with addition of a linear combination of 2-solution coupling. (a) $y_1 \times y$; (b) $y_2 \times y$; (c) $y_s = \frac{(y_1 +y_2)}{2} \times y$. Left: synchronization analysis; Right: temporal series.}
\end{figure}

\begin{figure}[htb]
\centering 
 \includegraphics[width = 8.0 cm]{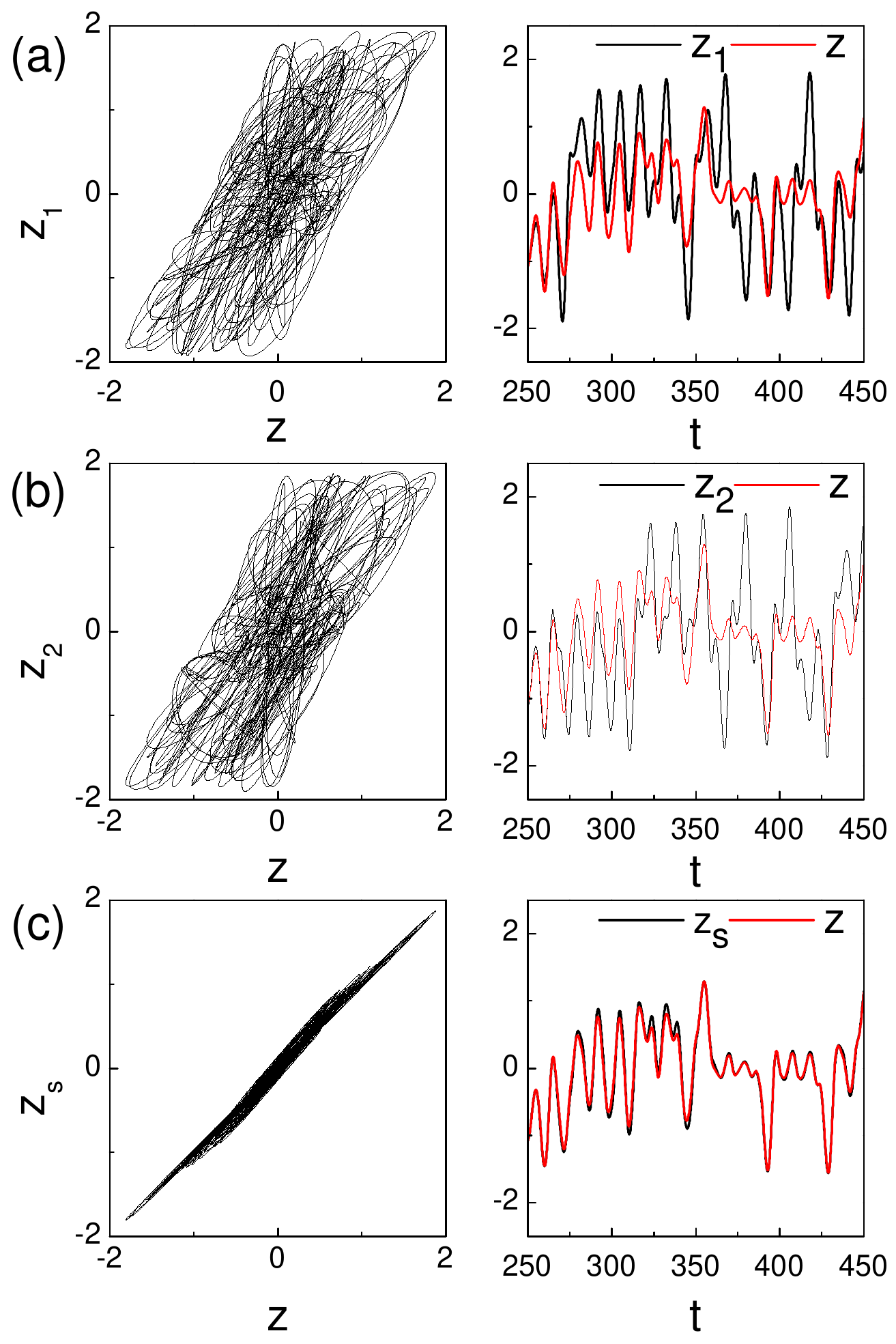}
 \caption{\label{fig3} Output of the Gauthier-Bienfang nonlinear system with addition of a linear combination of 2-solution coupling. (a) $z_1 \times z$; (b) $z_2 \times z$; (c) $z_s = \frac{(z_1 +z_2)}{2} \times z$. Left: synchronization analysis; Right: temporal series.}
\end{figure}

To analyze the synchronization, the components $(x,y,z)$ of the response solution are compared to each variable $(x_n,y_n,z_n)$, with $n = 1,2$, of the individual drive oscillators (see Figures \ref{fig1}-\ref{fig3}). 
We show in Figure \ref{fig1}(a,b) the graphs $x \times x_1$ and $x \times x_2$, as well as the temporal series for $x$, $x_1$ and $x_2$. Clearly, the system does not synchronize with any of the two individual components, with both phase and amplitude uncorrelated. These comparative trajectories and temporal series are shown in Figure \ref{fig1}(c) for the component $x_s = (x_1+ x_2 )/2$ of the sum of the two drive variables $x_1$ and $x_2$. The component $x$ of the response oscillator synchronizes with the component $x_s$ of the weighted linear sum signal. 

Figures \ref{fig2} and \ref{fig3} exhibit graphs $y \times y_{n}$ and $z \times z_{n}$ for $n = 1,2$ and temporal series for the components of the drive and receiving systems. The strong synchronization between $\bf{X}$ and $\bf{X_s}$ is displayed in all the components synchronization curves, $\bf{X} \times \bf{X_s}$, as well as in the temporal series $\bf{X}(t)$ and $\bf{X_s}(t)$. Comparing the solution variables $\mathbf{X}_n(t)$ $(n = 1,2)$ to those of the oscillator, $\bf{X}(t)$, we do not observe any regular relation of phase between the solution and individual variables (Figs. \ref{fig1}-\ref{fig3} (a,b)), but only with the sum of the solutions used to define $\bf{X_s}$(Figs. \ref{fig1}-\ref{fig3} (c)).

\begin{figure}[!h]
\centering 
 \includegraphics[width = 10.0 cm]{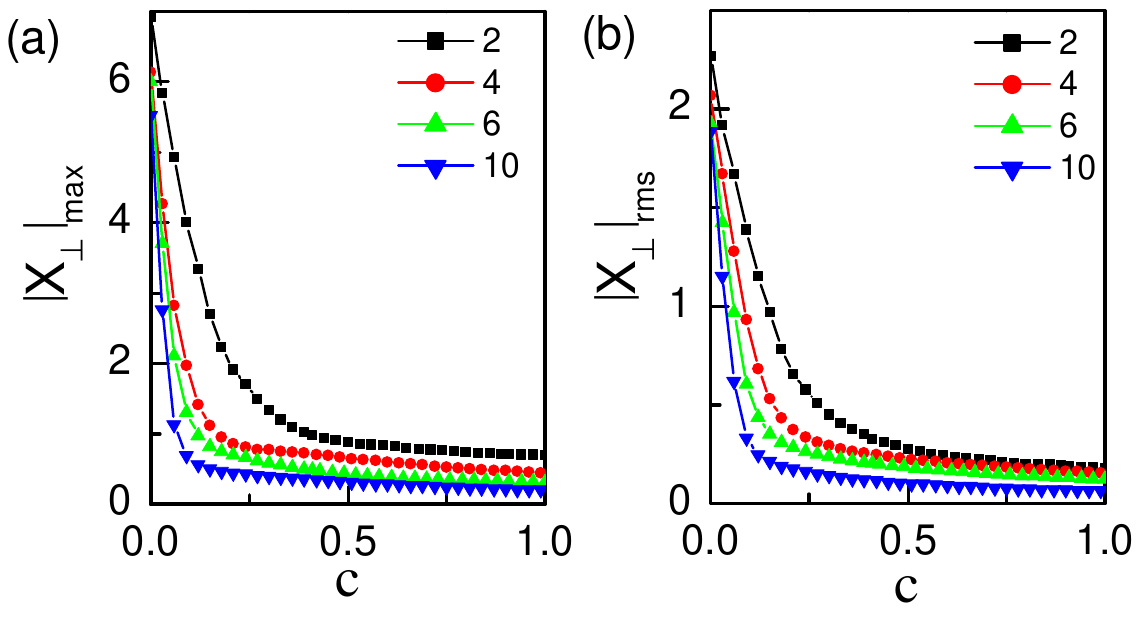}
 \caption{\label{fig4} Measure of the convergence of the oscillator trajectory to a sum signal. (a) $|\mathbf{X}_{\bot}|_{max}$ and (b) $| \mathbf{X}_{\bot} |_{rms}$,  for the sum of: 2 (black squares),  4 (red circles), 6 (green up triangle), and 10 (blue down triangle) solutions.}
\end{figure}

We have investigated the behaviour of the solution of equation (\ref{eq2}) as a function of the number $N$ of solutions of Eq.(\ref{eq1}) composing $\bf{X}_s$. For simplification purposes we consider $c_n = c$ $(n = 1, 2,....N)$. We define the vector $\bf{X}_{\bot} = (\bf{X}_s - \bf{X})$ and show in Figure \ref{fig4} the behavior of the distance $|\bf{X}_{\bot}|$ \cite{modulus} as a function of the coupling coefficient $c$, that evidences the convergence to synchronization for a large number of added solutions. Figure \ref{fig4}(a) exhibits   $|\mathbf{X}_{\bot} |_{max}$, the maximum value of $|\mathbf{X}_{\bot} |$ in the temporal series, which is more sensitive to local instability, and Figure 4(b) displays $| \mathbf{X}_{\bot} |_{rms}$, the average value of $|\mathbf{X}_{\bot}|$ in the temporal series, as a measure of the global stability of solution $\bf{X}$. A monotonic convergence of $|\mathbf{X}_{\bot}|$ to zero is observed as the number of solutions added into $\bf{X}_s$ is increased.

The decreasing of both $|\mathbf{X}_{\bot}|_{max}$ and  $| \mathbf{X}_{\bot}|_{rms}$ as $N$ increases confirms the behaviour observed in the graphs $\frac{1}{2}(\mathbf{X}_1+ \mathbf{X}_2) \times \bf{X}$ of Figs \ref{fig1}-\ref{fig3}(c) in the case of two solutions only: the convergence of $\bf{X}$ toward $\bf{X_s}$ improves as the number of solutions in $\bf{X_s}$ increases. We have likewise applied this technique to Lorenz and R\"{o}ssler systems. Both also converge to the solution-sum, yet R\"{o}ssler shows a slower convergence for the $z$-variable with the number of solutions.
\newpage

\section{Conclusion}

We showed that is possible to create a state of coherence in a classical, chaotic system. The non-linear system oscillates, following trajectories that are not a simple solution of the system equations but a \textit{linear} combination of solutions. The increased sensitivity of nonlinear systems to a high number of solutions, even when only a partial information is transmitted through one or few variables, gives an insight into  processes where a large number of inputs determine a single output, as occurs in complex networks such as neural systems.

\begin{acknowledgments}
This work was supported by Universidade Federal Rural de Pernambuco (UFRPE) and the Brazilian agencies  CNPq/Universal, CAPES, and FACEPE.
\end{acknowledgments}
%
%
\bibliography{References}

\begin{thebibliography}{5}
\bibitem [{mod()}]{modulus}%
  \BibitemOpen
  \href@noop {} {}\bibinfo {note} {The distance $\vert \mathbf{X}_{\bot}\vert$
  is defined as $\vert \mathbf{X}_{\bot}\vert \equiv (\vert x_s - x \vert +
  \vert y_s - y\vert) + \vert z_s - z \vert)$}\BibitemShut {NoStop}%
\end{thebibliography}
\end{document}